**Any quantum state can be cloned in the presence of closed timelike curves**

D. Ahn[1*], T. C. Ralph[2] and R. B. Mann[3]

[1] *Center Quantum Information Processing, Department of Electrical and Computer Engineering, University of Seoul, Seoul 130-743, Republic of Korea*

[2] *Center for Quantum Computer Technology, Department of Physics, University of Queensland, Brisbane 4072, QLD, Australia*

[3] *Department of Physics and Astronomy, University of Waterloo, Waterloo, Ontario, N2L, 3G1, Canada*

**Abstract**

The possible existence of closed timelike curves (CTCs) draws attention to fundamental questions about what is physically possible and what is not. An example is the "no cloning theorem" in quantum mechanics, which states that no physical means exists by which an unknown arbitrary quantum state can be reproduced or copied perfectly. Using the Deutsch approach, we show here that this theorem can be circumvented in the presence of closed timelike curves, allowing the cloning of an unknown arbitrary quantum state chosen from a finite alphabet of states. Since the "no cloning theorem" has played a central role in the development of quantum information science, it is clear that the existence of CTCs would radically change the rules for quantum information technology. Nevertheless we show that this type of cloning does not violate no-signalling criteria.

*To whom correspondence should be addressed. E-mail: dahn@uos.ac.kr

The possible existence of closed timelike curves (CTCs) allowing time travel draws attention to fundamental questions about what is physically possible and what is not [1-11]. An example is the impossibility theorem in quantum mechanics called the "no cloning theorem" [12-14], which states that there exists no physical means by which an unknown arbitrary quantum state can be reproduced or copied if chronology is respected. Recently, Brun et al. [10] have conjectured a CTC-assisted cloning with fidelity approaching one at the cost of increasing the available dimensions in ancillary and CTC resources. We show here explicitly that in the presence of closed timelike curves quantum mechanics allows the cloning of an unknown arbitrary quantum state with both finite dimensional ancillary and CTC resources in a succinct way. One of the original arguments against cloning was that it would allow signaling, i.e. faster than light communication, when applied to an entangled state [12]. We also show that this type of cloning does not violate no-signaling criteria. Should the ability to manipulate closed timelike curves ever become possible, our research suggests that new possibilities in quantum information technology would emerge, including eavesdropping without detection and perfect quantum broadcasting.

The most widely accepted model for calculating the evolution of a quantum system in the presence of closed timelike curves, proposed by Deutsch [3] and Politzer [4], involves a self-consistent solution for the density matrix. In this model, a unitary interaction U of a chronology respecting (CR) quantum system with a quantum system traveling around the closed timelike curve (CTC) leads to self-consistent evolution of an initial data which does not give rise to any of the typical "patricidal paradoxes" usually associated with time travel[3,4]. As Deutsch's solution relies only on the geometry of spacetime described by general relativistic closed timelike curves, we refer to the CTCs in our study as "geometric closed timelike curves" The quantum systems are the density matrices of quantum mechanics and the dynamics are augmented from the usual linear evolution. For each input density matrix $\rho_{CR}$, the CTC quantum system is postulated to find at least one fixed-point $\rho_{CTC}$ such that

$$\rho_{CTC} = Tr_{CR}\left(U \rho_{CR} \otimes \rho_{CTC} U^{\dagger}\right), \qquad (1)$$

which is called a self-consistency condition for the CTC system [3, 4]. The final state of the CR system is then defined in terms of the fixed point as [3,4]

$$\rho'_{CR} = Tr_{CTC}\left(U \rho_{CR} \otimes \rho_{CTC} U^{\dagger}\right). \qquad (2)$$

The induced mapping $\rho_{CR} \to \rho'_{CR}$ is nonlinear because the fixed point $\rho_{CTC}$ also depends on the input state $\rho_{CR}$. It is this nonlinearity that would distinguish the CTC system from ordinary quantum mechanics.

It is an interesting question from a fundamental physics point of view whether operations forbidden by the linearity of quantum mechanics would be permissible in the presence of CTC systems. Previously, it has been argued that the CTC nonliearity could improve quantum state discrimination [10] or speed up hard computations [8]. An alternative viewpoint that such increased power is not implied by CTCs [11] has been argued to be inconsistent with the Deutsch model [15,16]. It was discovered by Wooters and Zurek [12] almost three decades ago that the linearity of quantum mechanics leads to an impossibility theorem called the "no cloning theorem". The theorem dictates that no apparatus exists that will copy an arbitrary quantum state. It does not rule out the possibility of copying orthonormal states by a device designed especially for that purpose, but it does rule out the existence of a device capable of cloning an arbitrary state.

In this Letter, we show that an apparatus exists that will clone an arbitrary quantum state chosen from a finite alphabet of states in the presence of a closed timelike curve. The general problem, posed formally, is as follows: A CR quantum system AB is composed of two parts, A and B, each belonging to an N dimensional Hilbert space. System A is prepared in one state from a set $A = \{\rho_j\}_{j=0}^{N-1}$ of N quantum states. System B, slated to receive the unknown state, is in a standard quantum state $\Sigma$. The initial state of the composite CR system AB is in the product state $\rho_s \otimes \Sigma$, where $s = 0,1,\cdots N-1$ specifies which state is to be cloned. We ask whether there is any physical process that leads to an evolution of the form

$$Tr_{CTC}\left(U\rho_s \otimes \Sigma \otimes \rho_{CTC}^s U^\dagger\right) = \rho_s \otimes \rho_s \tag{3}$$

for some unitary operator U and a fixed point $\rho_{CTC}^s$ which satisfies a self-consistency condition for CTC system

$$\rho_{CTC}^s = Tr_{AB}\left(U\rho_s \otimes \Sigma \otimes \rho_{CTC}^s U^\dagger\right) \tag{4}$$

for each *s*. To demonstrate how to circumvent the no-cloning theorem, we employ the concept of fidelity $F(\rho_i, \rho_j)$ between two density operators, defined by [14,17]

$$F(\rho_i, \rho_j) = Tr\sqrt{\rho_i^{1/2} \rho_j \rho_i^{1/2}} \tag{5}$$

where for any positive operator $O$, $O^{1/2}$ denotes its unique positive square root. Fidelity is an analog of the modulus of the inner product for pure states and can be interpreted as a measure of distinguishability for quantum states: it ranges between 0 and 1, reaching 0 if and only if the states are orthogonal and reaching 1 if and only if $\rho_i = \rho_{j1}$. It is invariant under the interchange $i \leftrightarrow j$ and under the unitary transformation $\rho_s \to U\rho_s U^\dagger$ for any unitary transformation U [17]. Also, from the properties of the direct product, we have [17]

$$F(\rho_i \otimes \sigma_i, \rho_j \otimes \sigma_j) = F(\rho_i, \rho_j) F(\sigma_i, \sigma_j). \tag{6}$$

Furthermore, if $\sigma = Tr_C(\tilde{\sigma})$ and $\tau = Tr_C(\tilde{\tau})$ where $Tr_C$ denotes partial trace over the subsystem $C$, we have [14,17]

$$F(\tilde{\sigma}, \tilde{\tau}) \leq F(\sigma, \tau) \tag{7}$$

referred to as the partial trace property. Equality holds when there is an optimal positive operator-valued measure (POVM) [14,17].

When there is no CTC system interacting with the CR system AB, then the cloning condition is simplified such that

$$Tr_C\left(U\rho_s \otimes \Sigma \otimes Y U^\dagger\right) = \rho_s \otimes \rho_s \tag{8}$$

where C is an auxiliary quantum system in some standard state $Y$. In this case, it can be shown that the optimal POVM exists and from equations (6) to (8), we obtain

$$F(\rho_i, \rho_j) = F\left(Tr_C\left(U\rho_i \otimes \Sigma \otimes Y U^\dagger\right), Tr_C\left(U\rho_j \otimes \Sigma \otimes Y U^\dagger\right)\right) = F(\rho_i \otimes \rho_i, \rho_j \otimes \rho_j) = F(\rho_i, \rho_j)^2$$

which means that $F(\rho_i, \rho_j) = 1$ or 0; i.e., $\rho_i$ and $\rho_j$ are identical or orthogonal. As a result, there can be no cloning for density operators with nontrivial fidelity when there is no violation of chronology [14].

On the other hand when the CR system AB is interacting with the CTC system, from the properties of the direct product, we have

$$F\left(U\rho_i \otimes \Sigma \otimes \rho_{CTC}^i U^\dagger, U\rho_j \otimes \Sigma \otimes \rho_{CTC}^j U^\dagger\right) = F(\rho_i, \rho_j) F(\rho_{CTC}^i, \rho_{CTC}^j). \tag{9}$$

Assuming (3) and making use of (4) and (7), we have the following partial trace properties for the CTC system:

$$F(\rho_i,\rho_j)F(\rho^i_{CTC},\rho^j_{CTC}) \le F(\rho_i,\rho_j)^2, \tag{10}$$

and $$F(\rho_i,\rho_j)F(\rho^i_{CTC},\rho^j_{CTC}) \le F(\rho^i_{CTC},\rho^j_{CTC}) \text{ for } i \ne j. \tag{11}$$

Due to the requirement of different fixed points $\rho^i_{CTC}$ and $\rho^j_{CTC}$, the existence of an optimal POVM for equalities in equations (10) and (11) is not guaranteed. From equation (10), we have $F(\rho_i,\rho_j) \ge 0$ or $F(\rho^i_{CTC},\rho^j_{CTC}) \le F(\rho_i,\rho_j)$ for non trivial fidelity for cloning of density operators when the CR quantum system is interacting with the CTC system. Above results show that as long as the density matrices in the CTC and CR systems satisfy the condition $F(\rho^i_{CTC},\rho^j_{CTC}) \le F(\rho_i,\rho_j)$, any state specified as an initial data for the CR system can be copied faithfully.

As an example, consider a set $\{|\psi_j\rangle\}_{j=0}^{N-1}$ of $N$ distinct states in a space of dimension $N$. The set $\{|\psi_j\rangle\}$ is not necessarily an orthonormal set. It can be shown [7] that there is a unitary transformation $U_j$ such that $U_j|\psi_j\rangle = |j\rangle$ where the states $|j\rangle$ are a standard orthonormal basis for the $N$-dimensional Hilbert space. We now construct a CTC containing an $N$-dimensional system in the loop. We prepare the input system A consisting of one of the states $|\psi_j\rangle$. The input system B is prepared as $\Sigma = |0\rangle\langle 0|$. The evolution operator $U$ for the total system in the presence of a CTC is given by $U = T_2 T_1 SVW$ where $W = SWAP(A \leftrightarrow CTC)$, $V = CSUM \otimes I_{CTC}$, $S = I_A \otimes \sum_k |k\rangle\langle k| \otimes U_k$, $T_1 = \sum_l |l\rangle\langle l| \otimes U_l^\dagger \otimes I_{CTC}$ and $T_2 = \sum_m U_m^\dagger \otimes I_B \otimes |m\rangle\langle m|$. Here, $CSUM$ acts on orthonormal basis according to $CSUM(|i\rangle \otimes |j\rangle) = |i\rangle \otimes |j+i(\bmod N)\rangle$ [18]. Before the interaction, the CTC system is in the state $\rho_{CTC}$ and the chronology respecting system AB is in the state $\rho_{CR} = |\psi_j\rangle\langle\psi_j| \otimes \Sigma$.

It is straightforward to show that the solution $\rho_{CTC} = |j\rangle\langle j|$ uniquely satisfies the self-consistency condition given by equation (2). The output state of the chronology respecting system is given by

$$\rho'_{CR} = Tr_{CTC}\left(T_2 T_1 S V W |\psi_j\rangle\langle\psi_j| \otimes \Sigma \otimes |j\rangle\langle j| W^\dagger V^\dagger S^\dagger T_1^\dagger T_2^\dagger\right)$$
$$= \left(U_j^\dagger |j\rangle\langle j| U_j\right) \otimes \left(U_j^\dagger |j\rangle\langle j| U_j\right) = |\psi_j\rangle\langle\psi_j| \otimes |\psi_j\rangle\langle\psi_j|,$$
(12)

which shows that the CTC system indeed allows the cloning of arbitrary pure quantum states. It is clear that the above solution satisfies the cloning condition $F(\rho^i_{CTC}, \rho^j_{CTC}) \leq F(\rho_i, \rho_j)$ because $\rho^j_{CTC}$ are orthogonal and $F(\rho^i_{CTC}, \rho^j_{CTC}) = 0$ for $i \neq j$. This is an example of perfect broadcasting in which the density operator of each of the separate systems is the same as the state to be broadcast [14].

As a second example, consider a mixed state $\rho_A = \sum_j \lambda_j |j\rangle\langle j|$ for the input system A. The input system B is prepared as $\Sigma = |0\rangle\langle 0|$ as before. Here $\{|j\rangle\}$ is an orthonormal basis which makes $\rho_A$ diagonal. The evolution operator $U$ for the total system in the presence of a CTC is given by $U = V W_1 W_2$ where $W_1 = SWAP(A \leftrightarrow CTC)$, $W_2 = I_A \otimes SWAP(B \leftrightarrow CTC)$ and $V = I_A \otimes CSUM$. Then it is straightforward to see that $\rho_{CTC} = \sum_k \lambda_k |k\rangle\langle k|$ and

$$Tr_{CTC}\left(U \rho_A \otimes |0\rangle\langle 0| \otimes \rho_{CTC} U^\dagger\right) = \left(\sum_j \lambda_j |j\rangle\langle j|\right) \otimes \left(\sum_l \lambda_l |l\rangle\langle l|\right).$$
(13)

One of the original arguments against cloning was that it would allow signaling, i.e. faster than light communication, when applied to an entangled state. Let's assume that one party of the entangled state, say Alice is locally interacting with the CTC for the cloning. For example, if the state vector of the entangled state is given by

$$|\Psi\rangle_{AR} = \frac{1}{\sqrt{2}}\left(|0\rangle_A |1\rangle_R + |1\rangle_A |0\rangle_R\right),$$

then the output of the chronology respecting system would be

$$\rho_{tot} = Tr_{CTC}\left(U |\Psi\rangle_{AR}\langle\Psi| \otimes |\Sigma\rangle\langle\Sigma| \otimes \rho_{CTC} U^\dagger\right).$$
(14)

Here the unitary operator $U$ is not acting on the state belonging to Rob. By taking the partial trace respect to Rob's state and from equations (3), (12) and (13), we get

$$\begin{aligned} Tr_R(\rho_{tot}) &= Tr_{CTC}\left(U\ Tr_R(|\Psi\rangle_{AR}\langle\Psi|)\otimes|\Sigma\rangle\langle\Sigma|\otimes\rho_{CTC}U^\dagger\right) \\ &= Tr_{CTC}\left(U\rho_A\otimes|\Sigma\rangle\langle\Sigma|\otimes\rho_{CTC}U^\dagger\right) \\ &= \rho_A\otimes\rho_A \end{aligned} \qquad (15)$$

where $\rho_A = Tr_R(|\Psi\rangle_{AR}\langle\Psi|)$.

Since the clone is the reduced density operator of the initial entangled state $|\Psi\rangle_{AB}$, no correlations remain between the clone and the other half of the entangled state. This clearly denotes that faster than light communication does not result from this type of cloning.

Our results have further implications. In quantum cryptography, the legitimate users of a communication channel encode the bits 0 and 1 into nonorthogonal pure states to ensure that any eavesdropping is detectable since eavesdropping necessarily disturbs the state sent to the legitimate user due to the no-cloning theorem. If nature allows CTC's, an eavesdropping party with access to a CTC can prepare the ancillary state $\Sigma$ and obtain a perfect copy of the input state initially possessed by the system A. However entanglement-based QKD (quantum key distribution) would remain secure against the type of cloning we described in this work because there is no correlation as shown by equation (15).

From an historical point of view, many insights obtained from the analysis of thought experiments that might be impossible to actually realize contributed significantly to the development of quantum mechanics [3]. Investigations of quantum mechanics in the presence of the closed timelike curves, even if they remain only theoretical constructs, may well contribute the development of a yet unknown full theory of quantum gravity.

**Acknowledgements**
D. A was supported by the University of Seoul through the University Research Grant 2008. R. B. M was supported by the Natural Sciences and Engineering Research Council of Canada.

**References**

1. M. S. Morris, K. S. Thorne, and U. Yurtserver, Phys. Rev. Lett. 61, 1446 (1988).
2. J. R. Gott III, Phys. Rev. Lett. 66, 1126 (1991).
3. D. Deutsch, Phys. Rev. D. 44, 3197 (1991).
4. H. D. Politzer, Phys. Rev. D 49, 3981 (1994).
5. J. B. Hartle, Phys. Rev. D. 49, 6543 (1994).
6. M. J. Cassidy, Phys. Rev. D. 52, 5676 (1995)
7. S. W. Hawking, Phys. Rev. D 52, 5681 (1995).
8. D. Bacon, Phys. Rev. A 70, 032309 (2004).
9. T. C. Ralph, Phys. Rev. A 76, 012336 (2007).
10. T. A. Brun, J. Harrington, and M. M. Wilde, Phys. Rev. Lett. 102, 210402 (2009).
11. C. H. Bennett, D. Leung, G. Smith, and J. A. Smolin, Phys. Rev. Lett. 103, 170502 (2009).
12. W. K. Wooters, W. H. Zurek, Nature 299, 802 (1982).
13. D. Dieks, Phys. Lett. 92A, 271 (1982).
14. H. Barnum, C. M. Caves, C. A. Fuchs, R. Jozsa, and B. Schumacher, Phys. Rev. Lett. 76, 2818 (1996).
15. T. C. Ralph, C. R. Myers, arXiv:1003.1987 (2010).
16. E. G. Cavalcanti, N. C. Menicucci, arXiv:1004.1219 (2010).
17. R. Jozsa, J. Modern Opt. 41, 2315 (1994).
18. D. Gottesman and J. Prekill, J. High Energy Phys. 03 (2004) 026